\documentclass[aps,twocolumn,superscriptaddress,preprintnumbers]{revtex4}

\usepackage{epsfig}
\usepackage{dcolumn}
\begin{document}
\newcommand{\fig}[2]{\includegraphics[width=#1]{#2}}

\title{Fractional quantum Hall effect in the absence of Landau levels}
\author{D. N. Sheng}
\affiliation{Department of Physics and Astronomy, California State
University, Northridge, CA 91330,USA}
\author{Zheng-Cheng Gu}
\affiliation{ Kavli Institute for Theoretical Physics, University of
California, Santa Barbara, CA 93106, USA}
\author{Kai Sun}
\affiliation{ Condensed Matter Theory Center and Joint Quantum
Institute, Department of Physics, University of Maryland, College
Park, MD 20742, USA }
\author{L. Sheng\footnote{email: shengli@nju.edu.cn}}
\affiliation{National Laboratory of Solid State Microstructures and
Department of Physics, Nanjing University, Nanjing 210093, P. R. China}

\date{\today}
\def\PRL#1{{\sl Phys. Rev. Lett. }{\bf #1}}
\def\PRB#1{{\sl Phys. Rev.  B }{\bf #1}}
\begin{abstract}
It has been well-known that topological phenomena with
fractional excitations, i.e., the fractional quantum Hall effect
(FQHE)~\cite{Tsui1982} will emerge when electrons move in Landau
levels. In this letter, we report  the discovery of 
the FQHE in the absence of Landau levels
in an interacting fermion model. The non-interacting part of our
Hamiltonian is the recently proposed topologically nontrivial flat
band model on the checkerboard lattice~\cite{sun}. In the presence
of nearest-neighboring repulsion ($U$), we find that at $1/3$
filling, the Fermi-liquid state  is unstable towards FQHE. At $1/5$
filling, however, a next-nearest-neighboring repulsion is needed for
the occurrence of the $1/5$ FQHE when $U$ is not too strong. We
demonstrate the characteristic features of these novel states and
determine the phase diagram correspondingly.
\end{abstract}
\maketitle

As one of the most significant discoveries in modern condensed
matter physics, FQHE~\cite{Tsui1982}  has attracted intense
theoretical and experimental studies in the past three decades.
The wavefunctions proposed by Laughlin\cite{Laughlin1983,Haldane1983,Halperin1984} first explained FQHE by introducing fractionally charged quasi-particles. Later, the ideas
of flux attachment and the composite Fermi-liquid theory\cite{Jain1989}
have provided us a rather simple and deep
understanding for the nature of the FQHE.
Among many of its interesting and unique properties, two of the most striking features
of FQHE are: (1) fractionalization,  where quasi-particle
excitations carry fractional quantum numbers and
even fractional statistics of the constituent particles,
and (2) topological degeneracy, where the number of degenerate ground
states  responds nontrivially to the changing of the topology of the
underlying
manifold. These two phenomena are the central ideas of the topological ordering~\cite{Wen1990}. In addition,
on the potential application side, they are also the key components in the studies
of topological quantum computation~\cite{Nayak2008}.

Since 1988,  great efforts have been made to study 
quantum Hall effects in lattice models without Landau 
levels\cite{Haldane1988}. The first
such an example is the theoretical model proposed by
Haldane~\cite{Haldane1988}, where it was demonstrated that in a
half-filled honeycomb lattice, an integer quantum Hall state can be
stabilized upon the introduction of imaginary hoppings. This study
was brought to the forefront again recently due to a major
breakthrough, in which a brand new class of topological states of
matter was discovered, known as the time-reversal invariant Z$_2$
topological insulators [See Refs.~\cite{Hasan2010,Qi2010} and
references therein]. From the topological point of the view, both
Haldane's model and Z$_2$ topological insulators can be considered
as generalizations of the integer quantum Hall effect. Opposite to
the FQHE, they don't support fractional excitations and the ground
states here have no topological degeneracy.

For fractional states, fractional Z$_2$ topological insulators are
found to be theoretically possible~\cite{Levin2009}.
However, it is highly unclear how to realize such kind of
states on lattice models.
This difficulty originates from the strong coupling nature of realizing fractionalized states.
In both the integer quantum Hall effect
and Z$_2$ topological insulators, all their essential topological
properties can be understood within a noninteracting picture.
However, for FQHE, interaction effects are expected to play the
vital role in stabilizing these fractionalized topological states. In
fact, without interactions, a fractional quantum Hall system will
become a  Fermi liquid
 due to the fractional filling factor.

Most recently, a series of models with topologically-nontrivial
nearly-flat band models have been proposed~\cite{Tang2010,
Neupert2010, sun}. 
These lattice models have topologically nontrivial bands, 
similar to Haldane's model and Z$_2$ topological insulators,
and their bandwidth can be tuned to much smaller than the band gap,
resulting in a nearly-flat band structure. In particular, based on
the mechanism of quadratic band touching~\cite{Tang2010, sun}, a
large class of flat band models have been explicitly obtained with
the ratio of the band gap over bandwidth reaching the high value of
$20-50$. Due to the strong analogy between these nearly-flat bands
and the Landau levels, it was conjectured that in these
models, FQHE (or
fractional topological insulators) can be stabilized in the presence
of repulsive interactions. However, the conjecture is challenged by
competing orders, e.g. the charge-density wave, and thus the fate of
these systems are unclear. More importantly, the flux attachment
picture~\cite{Jain1989} may also break down here and it is very
interesting to examine the nature of the corresponding emergent
fractionalized quantum state if it can be realized in such a model
without Landau levels.

\begin{figure}[h]
{\includegraphics[width=0.45\textwidth]{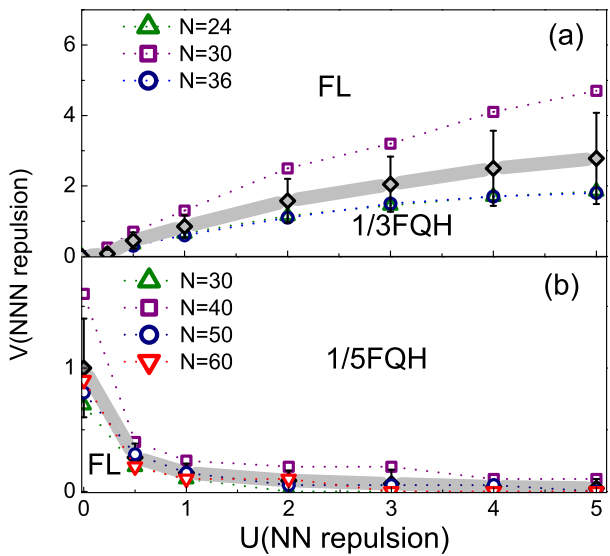}} \caption
{The phase diagram 
with particle filling factors (a) $1/3$ and (b) $1/5$ in the
parameter space of the nearest-neighbor ($U$) and next-
nearest-neighbor ($V$) interaction strengths for different system
sizes. In both Figs. (a) and (b), two phases are observed, the
Fermi-liquid phase (labeled as FL) and the fractional-quantum-Hall
phase (labeled as FQH). The dashed lines in both figures mark the
phase boundaries at different system size. The thick gray
lines mark the averaged phase boundary over different systems sizes
and the error bar mark the averaged deviations.}
\label{fig:fig1}
\end{figure}

In this Letter, based on exact calculations of finite-size systems,
we report the discovery of FQHE at filling factor $1/3$ and $1/5$
in models with topologically-nontrivial nearly-flat
bands~\cite{sun}.
The existence of the FQHE are confirmed by using two
independent methods, both of which are well-established and have
been widely adopted in the study for the traditional FQHE. We have
studied both the characteristic low-energy spectrum and the
topologically-invariant Chern
numbers~\cite{Thouless1982,Niu1985,Arovas1988,Huo1992,Sheng2003,Sheng2005a}
of the low-energy states. Here the first method detects directly the
topological degeneracy and the other is related with the phenomenon
of fractionalization.

We consider the following Hamiltonian on the checkerboard lattice:
\begin{eqnarray}
H=-H_{0}+U\sum_{\langle i, j \rangle} n_in_j+V\sum_{\langle\langle
i,j\rangle\rangle}n_in_j, \label{HAM}
\end{eqnarray}
where $H_0$ describe the short-range hoppings in the two-band checker-board-lattice
model defined in Ref.\cite{sun}.
 $n_i$ is the on-site fermion particle number operator.
The  nearest-neighboring (NN) and next-nearest-neighboring (NNN) bonds are represented by
$\langle i,j \rangle$ and  $\langle\langle i,j \rangle\rangle$,
respectively.

The effect of the NN $(U)$ and NNN $(V)$ interactions  
are summarized in the  phase diagrams Figs.\ 1(a) and (b).
At filling factor 1/3, the FQHE emerges  with the turn
on of a relatively small $U$.  Interestingly, the
FQHE phase remains robust at the large $U$ limit where particles
avoid each other at NN sites and can only be destroyed by an
intermediate $V$. At filling factor 1/5, the FQHE occurs in the most
region of the parameter space  as long as  the  NNN repulsion $V$
exceeds a critical value $V_c$, whose value drops to zero  for
larger $U$. The observed FQHE states are characterized by (i) nearly
$p$-fold ($p=3$ and $5$ for filling factors $1/3$ and $1/5$
respectively) degenerating ground states with the momentum quantum
numbers of these states related to each other by a unit momentum
translation of each particle as an emergent symmetry, (ii) a finite
spectrum gap separates the ground state manifold (GSM) from the
low-energy excited states with a magnitude dependent on the
interaction strengths $U$ and $V$, and (iii) the GSM carries a unit
total Chern number as a topological invariant protected by the
spectrum gap, resulting in a fractional effect of $1/p$ quantum for
each energy level in the GSM. We identify the quantum phase transition based on
the spectrum gap collapsing. As shown in Fig.\ 1, the Fermi-liquid
phase with gapless excitations is found for relatively strong NNN
interaction at 1/3 filling and for relatively weak NNN interaction
at 1/5 filling. 
This observation is consistent with Haldane's pseudopotential
theory\cite{Haldane1983} and previous studies on ordinary
FQHE, where a FQHE state is found to be sensitive to interactions
and other microscopic details
(e.g. the thickness of the 2D electron gas~\cite{Peterson2008}).

Here we present the studies of the low energy spectrum.
Consider a system of $N_x\times N_y$
unit cells  ($N_s=2N_xN_y$ sites) with twisted
boundary conditions: $T({\bf N}_j)|\Psi\rangle=e^{i\theta_j}|\Psi\rangle$,
where $T({\bf N}_j)$ is the  translation operator with $j=1,2$
representing the $x$ and $y$ directions, respectively. 
Note that the filling factor is
defined as the ratio of the number of particles ($N_p$) over
the number of unit cells ($N_x\times N_y$).
In the absence of impurities
the total momentum of  the many-body state is a conserved quantity and thus
the Hamiltonian can be diagonalized in each momentum sector for systems with
$N_s=24$ to $60$ sites (depending on filling factors). We consider periodic
boundary condition ($\theta_1=\theta_2=0$) first. Fig. 2(a)
illustrates the evolution of the low energy spectrum with changing
$U$ for $V=0$ and $N_s=2\times4\times6$  ($N_x=4$ and $N_y=6$)
at particle filling factor $\nu=1/3$. Here, the NN hopping strength
($t$)~\cite{sun} is set to unity.
We denote the momentum of a state  ${\bf q}=(2\pi k_x/N_x, 2\pi k_y/N_y)$ by using
two integers $(k_x, k_y)$ as shown in Fig. 2. For vanishing NN
interaction $U=0.0$ (the bottom panel of Fig.2(a)), the ground state
has $k=(0,0)$ while no particular structure is observed
in other $k$ sectors. For a weak interaction $U=0.2$, we find an interesting
change in the spectrum. There are two states with momenta $(0,k_y)$
($k_y=2$ and $4$), which have lowered their energies substantially.
 For a stronger interaction,  the energies of the three
states with $k_x=0$ and $k_y=0, 2,4$ 
form a nearly-degenerate GSM\cite{Wen1990} at $U>0.3$. In the mean time,
a sizable spectrum gap opens up, separating the GSM from the other excited
states as shown in the top panel of Fig.2(a) for
$U=1.0$.
The obtained three-fold ground state
near degeneracy and a robust spectrum gap are  the characteristic
features of the 1/3 FQHE phase, which emerge with the onset of the
NN repulsion $U$. By increasing the NNN repulsion $V$ to a certain
critical value $V_c$,
we have observed the collapsing of the spectrum gap, which
determines the boundary of the 1/3 FQHE phase, as shown in Fig.
1(a). Further evidence of the FQHE based upon
topological quantization will be presented later in Figs.\ 3.

\begin{figure}[h]
\begin{center}
\vskip 0.0cm \hspace*{-0.3cm} \fig{3.3in}{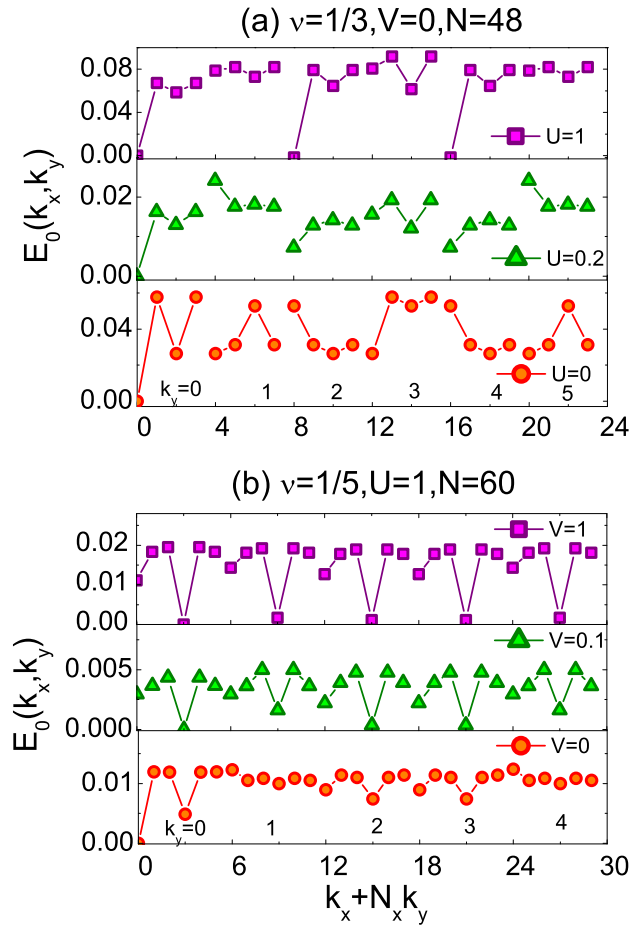} \caption {
The characteristic low-energy spectrum for (a) 1/3 filling  with
$V=0$, $U=0$ to $1.0$ ($U$ increases from bottom panel up)  and
$N_s=2\times4\times6$ sites; (b) 1/5 filling with $U=1$ and $V=0$ to
$V=1$ and $N_s=2\times6\times5$. We have shifted the ground state
energy to zero for comparison.  Nearly-degenerate GSM and the large
gap between GSM and excited states, observed in Fig. (a) at large
$U$ and Fig.(b) at large $V$ indicate the formation of the
fractional-quantum-Hall states. } \label{fig:fig2} \vskip -5cm
\end{center}
\end{figure}

In Fig. 2(b), we present the formation of the $1/5$ FQHE by
showing the energy spectrum at particle filling
factor $\nu=1/5$ with increasing $V$ at $U=1$ for a system with
$N_s=2\times6\times5$ ($N_x=6$ and $N_y=5$).
 From the bottom panel to the top panel,
 five states with momenta  $(2, k_y)$
 ($k_y=0,...,4$) form the nearly degenerate GSM with the increase
of $V$,
while a large spectrum gap is formed  at $V=1.0$.
The same feature of the energy spectrum is observed
for the whole regime of the 1/5 FQHE above
the critical $V_c$ line shown in
the phase diagram Fig. 1(b) while
$V_c$ drops to near zero for larger $U$.

We note that there is a small energy difference between the states
in the GSM in both filling factors.  This is a finite-size
effect\cite{tao,Sheng2003}  as
each of these states has to  fit into the lattice structure.
The finite-size effect
is substantially smaller for 1/5 FQHE comparing to the 1/3 case
due to the lower particle density.
Interestingly,  for all cases that we have checked for different
system sizes ($N_s=24-60$), the members of the GSM are always
related to each other through a momentum space translation as an
emerging symmetry of the system. Namely, if  $(k_1, k_2)$  is  the
momentum quantum number for a state in the GSM, then another state
in the GSM can be found in the momentum sector $(k_1+N_e,k_2+N_e)$
(modulo $(N_x, N_y)$).  This relation of the quantum numbers of the
GSM demonstrates the correlation between the real space and momentum
space in a manner precisely resembling  the FQHE in a uniform
magnetic field.
For the case where the particle number $N_p$ is integer multiples of
both $N_x$ and $N_y$ (e.g., $N_s=2\times3\times6$ and
$N_s=2\times5\times5$ at 1/3 and 1/5 fillings, respectively), all
the states of the GSM are indeed observed to fall into the same
momentum sector as expected.

\begin{figure}[h]
\begin{center}
\vskip-1.5cm \hspace*{-1.2cm} \fig{4.5in}{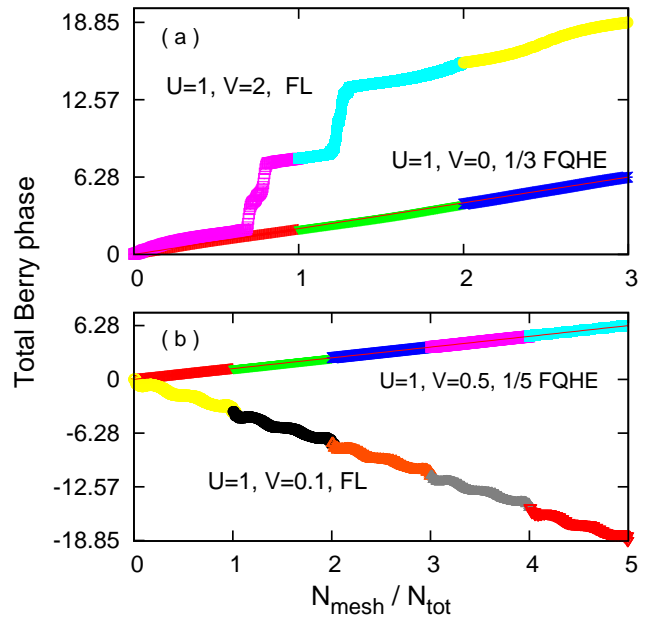}
\vskip-1cm \caption { The total Berry phase as a function of
$N_{mesh}/N_{tot}$, which measures the ratio of the area in the
boundary phase space for (a) 1/3 filling at $N_s=30$; (b) 1/5
filling at $N_s=40$.
In fractional-quantum-Hall phase,  a linear curve is observed whose slope is determined
by filling factor, as expected. In the Fermi liquid phase, we observed large fluctuation
and nonuniversal behaviors, indicating the absence of the
 topological quantization.}
\label{fig:fig3}
\vskip-1cm
\end{center}
\end{figure}

\begin{figure}[h]
\begin{center}
\vskip -0.5cm \hspace*{-1.0cm} \fig{4.5in}{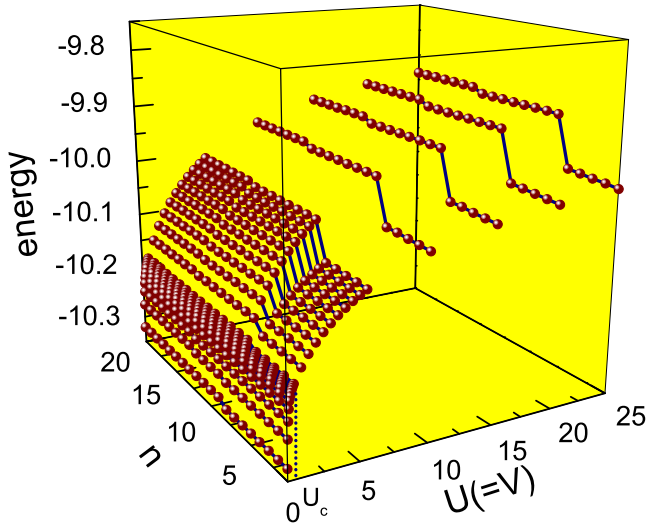} \caption{
Low energy spectrum for lowest $n$ states ($n=1,\ldots,20$) as a
function of $U$ (with $V=U$) for a system with $N_s=50$. The dashed
line marks the critical value $U_c$, at which the phase transition
between a Fermi-liquid phase and the FQHE takes place.}
\label{fig:3D} \vskip-1cm
\end{center}
\end{figure}

To further establish the existence of the FQHE here, we study the 
topological property of the
GSM through numerically inserting flux into the system using
the generalized boundary phases.
As first realized by Thouless and co-workers~\cite{Thouless1982,Niu1985,Arovas1988}, a topological
quantity of the wavefunction, known as the first Chern number,
distinguishes the quantum Hall states from other topological trivial
states. In particular, it has been known that one can detect
fractionalization phenomenon in FQHE by examining Chern number through
 inserting flux into the system~\cite{Huo1992,Sheng2003,Sheng2005a}.
The results of this calculation are shown in Fig.\ 3. With details presented in the Methods part,
 the total Berry phase as a function of $N_{mesh}/N_{tot}$ should be a linear function with slope
$2\pi/3$ and $2\pi/5$
for the FQHE with filling factors $1/3$ and $1/5$, respectively.
This agrees very well with the observation shown in Fig.\ 3.
On the other hand, for the Fermi-liquid phase,
strong fluctuations and nonuniversal behaviors of the Berry phase are found, suggesting the absence of  topological  quantization.

We further examine the phase at strong coupling limit for lower filling factor
$\nu=1/5$, where 1/5 FQHE demonstrates less sensitivity to either large $U$ or
$V$. We show the lowest 20 eigenvalues as a function of $U$ in Fig.
4, while we always set $V=U$ for simplicity. At small $U$, we see
the flatness of the spectrum  in consistent with a Fermi liquid phase with
small energy dispersion.  As we increase $U>U_c=0.35$, all the
lowest five states remain nearly degenerate, while higher
energy states jump a step up  making a robust gap between them and
the GSM. In fact, the 1/5 FQHE persists into infinite $U$ and $V$
limit as we have checked by projecting out the configurations with
double or more occupancy between a site and all its NN and NNN
sites. Physically, this can be understood as the particle at lower
filling has enough phase space within the lower Hubbard band, and
thus the FQHE remains intact. It would be very interesting to
establish a variational state for the FQHE on the flatband model,
which will be investigated in the future.

Finally, it is important to emphasize that the fractional
topological phases we found are very stable and the same effect
survives even if the hopping strengths are tuned by an amount of
$\sim 10\%$. On the experimental side, it is known that
checker-board lattice we studied can be found in condensed matter
systems (e.g. the thing films of
LiV$_2$O$_4$, MgTi$_2$O$_4$, Cd$_2$Re$_2$O$_7$, \emph{etc}), and the
imaginary hopping terms we required can be induced by
spin orbit coupling and spontaneous symmetry
breaking~\cite{Sun2009}. In addition, this lattice model also has
the potential to be realized in optical lattice system using
ultra-cold atomic gases, in which the tuning of the parameters are
much easier compared with condensed matter systems. Based on these
observations, we conclude that there is no fundamental challenge
preventing the experimental realization of these novel fractional
topological states, but further investigation is still needed in
order to discover the best experimental candidates.

\section*{Methods}

The Chern number of a many-body state can be obtained as:
\begin{eqnarray}
\label{integral}
C ={i\over 4\pi}\oint d\theta_j
\{\langle { \psi |{\partial \psi
 \over
\partial \theta_j}\rangle -
\langle {\partial \psi \over \partial \theta_j}|
\psi}
\rangle\},
\end{eqnarray}
where the closed path integral is along the boundary of a unit cell
$0 \leq \theta_j \leq 2\pi$ with $j=1$ and $2$, respectively.
The Chern number $C$  is also the Berry phase (in units of $2\pi$) accumulated for
such a state when the boundary phase evolves along the closed path.
Equation (2) can also be reformulated as an area integral over the unit cell
$C={1\over 2\pi}\int d\theta_1 d\theta_2  F$, where $F(\theta_1, \theta_2)$ is
the Berry curvature. To determine the Chern number accurately~\cite{Huo1992,Sheng2003,Sheng2005a},
we divide the boundary phase unit cell
into about $N_{tot}=200$ to $900$ meshes.
The curvature $F(\theta_1, \theta_2)$ is then given by the Berry phase of
each mesh divided by the area of the mesh.
The Chern number is
obtained by summing up the Berry phases of all the meshes.

We find that the curvature is in general  a very smooth function of
$\theta_j$ inside FQHE regime. For an example, the ground state
total Berry phase sums up to
$0.325\times2\pi$, slightly away from the 1/3 quantization for a
system with $N_s=30$, $U=1$ and $V=0$ at $1/3$ filling.
 Physically, as we start from one state with momentum
$(k_x, k_y)$ in the
GSM, it  evolves to another state with a different
momentum $k_x\rightarrow k_x+N_e$ ($k_y\rightarrow k_y+N_e$), when the boundary phase
along $x$ ($y$) direction is increased from $0$ to $2\pi$.  Thus,
with the insertion  of a flux, states evolve to each other within
the GSM. We observe that only the total Berry phase of the GSM
is precisely quantized to $2\pi$ and the total Chern number $C=1$
for all different choices of parameters
inside either the 1/3 or 1/5 FQHE regime of Fig. 1.

As we move across the phase boundary from the FQHE state
into the Fermi-liquid phase,
there is no well defined nearly-degenerate GSM or
spectrum gap, and the Berry curvature in general shows an order of
magnitude bigger fluctuations. The obtained total Chern integer
varies with system parameters (e.g., $U$ and $V$). In order to
illustrate this feature, we start from the lowest-energy eigenstate
and continuously increase the boundary phases for three periods,
which allows the first state to evolve into other states and
eventually return back to itself.
In Fig. 3, we plot the  accumulated total Berry phase as a function
of the ratio of the total meshes included $N_{mesh}$ over the total
number of meshes $N_{tot}$ in each period. For the system in the 1/3
FQHE phase with $N_s=30$, $U=1$ and $V=0$, the total Berry phase
follows a straight line in all three periods, well fitted by $\frac
{2\pi} 3N_{mesh}/N_{tot}$, indicating a nearly perfect linear law of
the  Berry phase to the area in the phase space with a deviation
around 10\%. In the Fermi liquid phase with $N_s=30$,  $U=1$ and $V=2$, we see
step-like jumps of the total Berry phase, with a magnitude in the
order of  $2\pi$, in sharp contrast to the linear law in the FQHE
phase.  The total Chern number for the Fermi-liquid state sums up to three,
indicating the decorrelation between three states. Different integer
values for the Chern number are found in this region (including
negative ones) with changing system parameters, demonstrating  a
measurable fluctuating Hall conductance if  particles are charged.
For 1/5 FQHE state, by following up
the Berry phase of five periods for the ground state as shown in Fig. 3(b),
we observe the same linear law with a slope of
$\frac {2\pi} 5$
and the total Chern number is quantized to one.
Interestingly, a negative integer Chern number for Fermi liquid for the
parameter $U=1$ and $V=0.1$ is observed confirming the nonuniversal
nature of the topological number for such a gapless system. We
conjecture that the Fermi liquid phase may be unstable towards Anderson
localization especially at lower filling factors, similarly to the
conventional FQHE systems\cite{Sheng2003}.

\section*{Acknowledgment} This work is  supported by
DOE Office of Basic Energy Sciences under grant
DE-FG02-06ER46305 (DNS), the NSF grant DMR-0958596 (for instrument),
the State Key
Program for Basic Researches of China under Grant Nos. 2009CB929504
and 2007CB925104, and the NSFC 10874066 and 11074110 (LS).
ZCG is supported in
part by the NSF Grant No. NSFPHY05-51164. KS acknowledges the support
from JQI-NSF-PFC,  AFOSR-MURI, ARO-DARPA-OLE, and ARO-MURI.


\end{document}